# QUANTIZED INTEREST RATE AT THE MONEY FOR AMERICAN OPTIONS

L. M. Dieng ( Department of Physics, CUNY/BCC, New York, New York)

**Abstract**: In this work, we expand the idea of Samuelson[3] and Shepp[2,5,6] for stock optimization using the Bachelier model [4] as our models for the stock price at the money ($X$[stock price]= $K$[strike price]) for the American call and put options [1]. At the money ($X= K$) for American options, the expected payoff of both the call and put options is zero. Shepp investigated several stochastic optimization problems using martingale and stopping time theories [2,5,6]. One of the problems he investigated was how to optimize the stock price using both the Black-Scholes (multiplicative) and the Bachelier (additive) models [7,6] for the American option above the strike price $K$ (exercise price) to a stopping point. In order to explore the non-relativistic quantum effect on the expected payoff for both the call and put options at the money, we assumed the stock price to undergo a stochastic process governed by the Bachelier (additive) model [4]. Further, using Ito calculus and martingale theory, we obtained a differential equation for the expected payoff for both the call and put options in terms of delta and gamma. We also obtained the solution to the non-relativistic Schroedinger equation as the expected payoff for both the call and put options. Then, we expressed the stochastic process that is the expected payoff for both the call and put options at the money in terms of the solution to the Schroedinger equation. For the first time, We showed the stochastic process that is the expected payoff at the money for both options to be an oscillatory function with quantized interest rates.

**INTRODUCTION**

Shepp [2,5,6] has solved several problems in stochastic optimization using martingale theory. He investigated stochastic optimization problems by assuming both the Black-Scholes (multiplicative) and Bachelier (additive) models as models for stock price movements. We will use as our model for the stock price movement the Bachelier model also known as the additive model [4]. Given below is the equation for the stock price movement:

$$X(t,\omega) = \mu + rt + \sigma W(t,\omega) \qquad 1$$

where at $t=0$, $X(0)=x$ is the initial stock price.

Equation (1) is called the Bachelier brownian model [8] with drift $\mu$ sometimes called the return on a given stock and volatility $\sigma$. It is important to mention here, both $\mu$ and $\sigma$ are assumed to be constant. Both parameters do not depend on time nor on the stock price. Equation (1) also satisfies the following stochastic Ito differential equation with constant parameters.

$$dX(t,\omega) = \mu dt + \sigma dW(t) \qquad 2$$

Traders on Wall Street prefer to price options using the Black-Scholes model, because the change of the stock price movement is proportional to the stock price itself which gives an exponential growth all the time. In contrast to the Black-Scholes model, the Bachelier model can have negative stock prices while in reality stock prices are non negative. However, in this work we are not interested in understanding single stock prices but we are interested in the probability distributions of stocks. We are currently working on a paper where the Black-Scholes model for stock price movement is used as our model.

Options are known within the financial market as a type of derivative, which are given to stock holders in order to hedge their positions against risky fluctuations of the stock price. They are divided into two categories: call and put options [1].

A call option is a contract that gives the right to the holder to buy the underlying asset at a lower stock price at the time of the contract, hoping that the stock price would go up above the strike price (exercise price) at some time before the expiration date. An investor holding such an option would wish the stock price to go higher than the strike price (or exercise price). Hence, the expected future payoff of the call option is given by:

$$e^{rt} \max(X(t)-K,0), \ X(t)>K \qquad 3$$

A put option is also a contract that gives the right to the holder to sell the underlying asset at a certain time when the stock price falls way below the strike price(exercise price). for a certain price (or strike price). An investor holding such a put option would wish the stock price to keep falling below the strike price (or exercise price). The expected future payoff of the put option is then given by:

$$e^{rt} \max(K-X(t),0), \ X(t)<K \qquad 4$$

where $X(t)$ is given by the Bachelier model (1) describing the stock price movements over a certain time period, $K$ is the strike price(exercise price) and $r$ is the risk free interest rate. We discounted equations (3) and (4) with the exponent $e^{rt}$ and in a non arbitrage situation, the expected return $\mu$ equals to the risk free interest rate $r$. Therefore replacing $\mu$ with $r$, the stochastic differential equation (2) becomes.

$$dX(t,\omega) = rdt + \sigma dW(t) \qquad 5$$

$X(t,\omega)$ is a stochastic process, $t \in T, \omega \in \Omega$ where $(\Omega, F, P)$ is a probability space on which there is a Wiener process, $W(t,\omega)$ measurable with respect to a family $F_t$ of $\sigma$ sub-fields $F$ is called a diffusion process if it satisfies the stochastic Ito differential equation (5) given above.

We will model the stock price movements in a risk free market environment with the assumption that both $\mu = r$ and $\sigma$ are constant with respect to the following parameters [1]: stock price, exercise price and time. It is important to mention here that, once an option is given; there are few parameters which could affect it during its life time and the parameters are: stock price (described with the Bachelier model [5]), risk free interest rate $r$, the strike price (exercise price) $K$, the volatility of the stock $\sigma$, time to maturity (expiration date of the option) $T$ and other dividends during the life time of the option if there are any to be paid [1]. One of the main assumptions in this work, is that there are no dividends to be paid during the entire life time of the options [1].

In this work, we will examine the American option, an American option is an option that can be exercised at any given time before the expiration time. Since an option is issued to the stock holder in order to hedge his or her position against risky fluctuations of the stock, therefore; he can get into an American call or an American put options contracts.

An American call option could be deep in the money when the expected payoff is always positive which is given by equation (3) or out of the money when the expected payoff is always negative which is given by equation (4). It could also be at the money when the expected payoff is equal to zero which is given by the following equation.

$$e^{rt} \max(X(t) - K, 0) = 0, \ X(t) = K \qquad 6$$

**Stopping Time or Markov Time and Martingale**

A random variable $\tau$ defined on the same sample space as a martingale or Markov process is a stopping time if for every time $t$, the event

$$\{\tau \leq t\} \in F_t$$

What does this really mean, it means that for any time $t$ you can tell whether or not $\tau \leq t$ that $\tau$ has occurred or not, by just knowing all the information up to time $t$, you don't need to have any information about the future.

A stochastic process $X_n(t,\omega)$, $n=0,1\ldots$ is called a martingale if the two conditions are met [9,10].

$$E[X_n] < \infty \quad (i)$$

$$E[X_{n+1}|X_0,\ldots\ldots X_n] = X_n \quad (ii)$$

What does this mean, (*i*) means the expected value of the price is always finite one and (*ii*) means that the expected value of the future price given the present price is equal to the present value of the price. Everything about the future price of the stock is incorporated into the present price, it also means there are no arbitrage opportunities.

In Figure 1, we have the three states that are when: The stock price is deep in the money (state a) with a positive payoff, the stock price is deep out of the money (state c) with a negative expected payoff and finally the stock price is at the money (state b) with zero expected payoff.

In this article, we will assume an American call option has been given to the stock holder which means; he can exercise the option at any given time before the expiration of the option. Hence, if the stock goes down deep out of the money below state b; He or She is not going to exercise because the expected payoff is negative. The best scenario here for the stock holder, is to wait until the stock goes up above the strike price $K$; then exercise with a positive payoff. From Figure 1, when the stock moves from deep out of the money (state c) toward the deep in the money (state b); it needs to go through state b where the expected payoff is zero, while the option holder is waiting for the stock price to go up.

We will investigate how market parameters such as the risk free interest rate, volatility of the stock behave with respect to each other while the stock holder is waiting for the stock price to rise deep in the money well above state b in Fig.1.

**Figure1**. Above at the money where $X=K$(state b) is called deep in the money($X>K$) and below is called deep out of the money($X<K$). At the money the expected payoff of the option is zero.

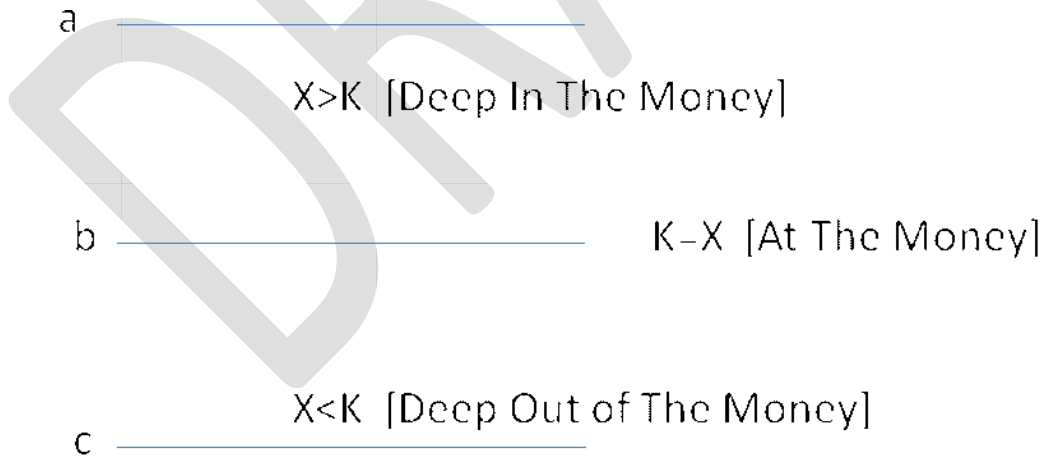

**Definition of supermartingale**

Since the expected payoff of the call option deep in the money is a positive function $V(X) > 0$ and hence it can be written in the following:

Using stopping time and Martingale theory, we wish the stock to reach the exercise price $K$, in order to do so the future payoff of the call option is defined to be the following:

$$V(X(t,\omega),K) = (X(t)-K,0)^+ \qquad 8$$

It is important to mention that, if an American put option were given to the stock holder, the expected payoff is the negative function $V(X) < 0$ and hence it can be written in the following:

$$V(X(t,\omega),K) = (K - X(t),0)^- \qquad 9$$

We will be dealing with equation (8), because we assumed the stock holder to have an American call option instead of an American put option. Hence, we want the future expected payoff to rise deep in the money; we could have also dealt with equation (9) if an American put option were given to the stock holder. But that is another different problem to be worked on.

Now, let's assume a stochastic function $Y(t,\omega)$ defined in terms of equation (8) and discounted at the risk free interest rate *r* via the exponential function. Then, the stochastic function is defined as:

$$Y(t,\omega) = V(X(t),K)e^{rt} \qquad 10$$

The stochastic function $Y(t,\omega)$ is nothing else but the future expected payoff discounted at the risk free interest rate of the stock holder. Since $Y(t,\omega)$ is defined in terms of equation (8), we will solve for *V(X(t), K)* in order to find its solution.

Using stopping time and martingale theory the future expected payoff of the American call option at the money (state b) in Fig.1 is zero.

Therefore if:

$\{\tau \leq t\} \in F_t$, there is a stopping time at which $\tau = t'$ where $\{t' < t\} \in F_t$

This corresponds to the following expected payoff at the money state b, where $X(\tau) = K$

$$V(X(\tau),K) = E[(X(\tau)-K,0)] \qquad 11$$

At $X(\tau) = K$, we want the sup of equation(11) to vanish that is:

$$V(X(\tau),K) = \sup_\tau E[(X(\tau)-K,0)]\big|_{X(\tau)=K} = 0$$

If *Y(t)* is a local supermartingale then it does have the following Ito differential equation:

$$dY(t) = \frac{\partial Y(t)}{\partial t}dt + \frac{\partial Y(t)}{\partial x}dx + \frac{1}{2}\frac{\partial^2 Y(t)}{\partial x^2}dx^2 + .. \qquad 12$$

in equation (12), the first, second and third terms correspond to the derivative of equation (10) with respect to time, the second and the third derivatives are with respect to the stock price respectively. Then, the conditional expectations on the Ito differential equation (12) given $F_t$ is defined by:

$$E[dY(t)|F_t] = 0 \qquad (iii)$$
$$E[dY(t)|F_t] \leq 0 \qquad (iv)$$

These expectations are called martingale and supermartingale respectively [8,9].
Taking the expectation of the differential equation (12) and substituting equation (10) into the differential, we will obtain the following:
Given that $Y(t)$ is defined by (10) and is a supermartingale, then we will obtain the following differential equation for $V(x)$:

$$E[(rVdt + V'dx + \frac{1}{2}V''(dx)^2)e^{rt}|F_t] \leq 0 \qquad 13$$

Since the stochastic differential equation is given by:

$$dX(t,\omega) = rdt + \sigma dW \Rightarrow (dX)^2 = (rdt + \sigma dW)^2$$

Substituting $dX$ and $(dX)^2$ into the expectation (13) we will obtain the differential equation

$$E[(rV + rV' + \frac{1}{2}\sigma^2 V'')e^{rt}dt|F_t] \leq 0 \qquad 14$$

since $Y(t)$ is a martingale then the expression inside the parenthesis in (14) should vanish, hence: $e^{rt} \neq 0$

$$\Rightarrow rV + rV' + \frac{1}{2}\sigma^2 V'' = 0 \qquad 15$$

Equation (15) is an ordinary differential equation with constant coefficients for the future payoff for both the call (put) option, where $r$ is interest rate and $\sigma^2$ is the volatility of the market. In the trader's language on Wall Street from equation (15), the first and second derivatives of the expected payoff $V$ with respect to the price $x$ is called delta and gamma respectively.

$$\Rightarrow rV + r\Delta + \frac{1}{2}\sigma^2 \Gamma = 0 \qquad 16$$

where $\Delta = \dfrac{\partial V(x,t)}{\partial x}$ and $\Gamma = \dfrac{\partial^2 V(x,t)}{\partial x^2}$

In the limit when $\Delta \to 0$, the second term in equation (16) will vanish and this corresponds to $\Delta$ hedging in finance. Hence, equation (16) can be rewritten in terms of gamma only with constant coefficients. Therefore, the quantum interpretation of the hedging is the potential energy of the price change which is zero.

$$\Rightarrow rV + \frac{1}{2}\sigma^2 \Gamma = 0 \qquad 17$$

We will treat equation (17) as the Shroedinger equation in quantum mechanics with constant interest rate and constant diffusion parameter. Rewriting equation (17), one will obtain the following:

$$\Rightarrow rV + D\Gamma = 0 \qquad 18$$

where the diffusion constant is found to be proportional to the volatility squared $D(\sigma) = \frac{1}{2}\sigma^2$.

We will solve equation (18) in terms of the constant diffusion constant $D(\sigma)$. By solving equation (18), we will obtain the dependence of interest rate with volatility and other market parameters and we will show that interest rate is quantized. It is important to mention here that interest rate, volatility and strike price (exercise price) are constant with respect to the price and time throughout this article.

We will seek the solution to equation (18) in the form:

$$V(x) = e^{\lambda x} \qquad 19$$

where $\lambda > 0$ and is constant. By substituting (19) and its second derivative with respect to the price $x$ into the differential equation (18), we will obtain the characteristic equation for $\lambda$

$$re^{\lambda x} + D\lambda^2 e^{\lambda x} = 0 \Rightarrow e^{\lambda x}(r + D\lambda^2) = 0$$

If $e^{\lambda x} \neq 0$ then:

$$r + D\lambda^2 = 0 \Rightarrow \lambda^2 = -\frac{r}{D}, \Rightarrow \lambda = \pm i\left(\frac{r}{D}\right)^{\frac{1}{2}}$$

Substituting $\lambda$ into (19) we will obtain the following oscillatory solution in terms of the price, interest rate and the diffusion constant. We found the expected future payoff for the America call option holder to be an oscillatory function with respect to the stock price.

$$V(x) = e^{\pm i\left(\frac{r}{D}\right)^{\frac{1}{2}}x} \Rightarrow V(x) = Ae^{i\left(\frac{r}{D}\right)^{\frac{1}{2}}x} + Be^{-i\left(\frac{r}{D}\right)^{\frac{1}{2}}x}, \text{ } A \text{ and } B \text{ are constant} \qquad 20$$

Now, we will express equation (10) in terms of the solution to the Schrodinger equation that is equation (20) to obtain the following stochastic expected future payoff at the money (state b) for the American call option;

$$Y(X,\omega) = V(x,t)e^{rt} = \left(Ae^{i\left(\frac{r}{D}\right)^{\frac{1}{2}}x} + Be^{-i\left(\frac{r}{D}\right)^{\frac{1}{2}}x}\right)e^{rt}$$

$$= (e^{rt})(A)\sin\left(\frac{r}{D}\right)^{\frac{1}{2}}x \qquad 21$$

We have found that the expected future payoff for the call (put) option at the money (state b in fig.1), is an oscillatory function with respect to interest rate, price and the diffusion constant $D(\sigma)$ and discounted with the exponent at the risk free interest rate $r$. In equation (21), $A$ is called the normalization constant and can be found by normalizing equation (20) as we do it in quantum mechanics.

At the money $X = K$ for the American option, the payoff of the option's holder should be zero and therefore we impose the following boundary condition for both the American call (put) option. Equation (23) could be set to zero if an American put option was given to the stock holder, which is not the case in this work. We assumed an American call option to have been given to the stock holder.

$$V(X(\tau), K) = \sup_\tau E[(X(\tau) - K, 0)]\big|_{X(\tau) = K} = 0 \quad\quad 22$$

$$V(X(\tau), K) = \sup_\tau E[(K - X(\tau), 0)]\big|_{X(\tau) = K} = 0 \quad\quad 23$$

In order to satisfy the boundary conditions on the expected future payoff at the money (state b), that is the expected future payoff should vanish at the money; equation (22) at the money should be equal to zero. Hence, one can apply the boundary condition at the money $X=K$ to equation (20) that is identical to equation (22):

$$V(X, r, \sigma) = A \sin\left(\frac{r}{D}\right)^{\frac{1}{2}} X \bigg|_{X(\tau)=K} = 0$$

$$\Rightarrow A \sin\left(\frac{r}{D}\right)^{\frac{1}{2}} K = 0 \text{, for } A \neq 0$$

$$\sin\left(\frac{r}{D}\right)^{\frac{1}{2}} K = 0 \Rightarrow \left(\frac{r}{D}\right)^{\frac{1}{2}} K = n\pi$$

Solving for $r$, the risk free interest rate in terms of the diffusion constant and the strike price (exercise price), we will obtain the following:

$$\left(\frac{r}{D}\right) K^2 = (n\pi)^2$$

$$\Rightarrow r = \left(\frac{D}{K^2}\right) n^2 \pi^2 \text{, where } n \text{ is an integer}$$

Or

$$r_n(\sigma, K, n) = \left(\frac{\sigma^2}{2K^2}\right) n^2 \pi^2 \quad\quad 24$$

We have obtained a time independent relationship between interest rate, volatility and exercise price at the money for the American call option. For a fixed exercise price $K$, we found the time-independent interest rate to be proportional to the volatility squared.

As we mentioned it above, one can find the constant *A* in (20) by normalizing the expected future payoff at the money that is:

$$\int V^*(x,r,\sigma,K)V(x,r,\sigma,K)dx = 1$$

$$\int_0^K A^2 \sin^2\left(\frac{r}{D}\right)^{\frac{1}{2}} x\,dx = 1 \Rightarrow A^2 = \frac{1}{\int_0^K \sin^2\left(\frac{r}{D}\right)^{\frac{1}{2}} x\,dx}$$

**Conclusion and Future Work**

We have assumed in this work that stock the holder has an American call option. We also assumed the stock to be deep out of the money, therefore the only option for the stock holder is to wait for the stock price to move deep out of the money (state c) towards the deep in the money (state a). With the assumptions above, the stock price sometimes is going to cross state b that is called at the money, where the expected payoff for the call option is zero. We were interested in finding out how market parameters vary while the stock holder with an American option is waiting for the stock price to get deep in the money. In order to find how market parameters such as interest rate, strike price (exercise price) and volatility behave in time when the stock price crosses state b at *X=K* where the expected payoff is zero. We used supermartingale and stopping time to obtain a stochastic Ito differential equation. We hedged the stochastic Ito differential equation in the limit when $\Delta \to 0$ at the money (state b) where the expected payoff is zero.

After hedging the Ito differential equation, we obtained a differential equation that we treated as the Schrodinger equation for a free particle in quantum mechanics with zero potential energy in the limit when $\Delta \to 0$. Then, we expressed the expected future stochastic payoff in terms of the future expected payoff of the call option.

We finally found the expected future stochastic payoff of the stock holder while waiting for the stock price to hit deep in the money domain well above state b but near state a to be an oscillatory function. We, at the end imposed a boundary condition when the stock price crosses at the money (state b) and found interest rate to be quantized in terms of the volatility and strike price. We initially were interested to show that the expected future stochastic payoff to be an oscillatory function with quantized interest rate.

In our future work, we'll assume our model for the stock price movement to be described with the Black-Sholes model. We'll numerically compute the expected payoff both the Bachelier and Black-Scholes models. We'll eventually compare the two models for research purposes.